\documentclass[reprint,superscriptaddress,showpacs,amsmath,amssymb,twocolumn]{revtex4-1}
\usepackage{graphicx}
\usepackage{dcolumn}
\usepackage{bm}
\usepackage{verbatim}
\usepackage{braket}

\begin{document}

\title{Polarimetry of photon echo on charged and neutral excitons\\ in CdTe/(Cd,Mg)Te quantum wells}

\author{S.~V.~Poltavtsev$^\dag$}
\email{sergei.poltavtcev@tu-dortmund.de}
\affiliation{Experimentelle Physik 2, Technische Universit\"at Dortmund, 44221 Dortmund, Germany}
\affiliation{Spin Optics Laboratory, St.~Petersburg State University, 198504 St.~Petersburg, Russia}
\author{Yu.~V.~Kapitonov$^\dag$}
\affiliation{Physics Faculty, St.~Petersburg State University, 199034 St.~Petersburg, Russia}
\thanks{These authors contributed equally.}
\author{I.~A.~Yugova}
\affiliation{Spin Optics Laboratory, St.~Petersburg State University, 198504 St.~Petersburg, Russia}
\author{I.~A.~Akimov}
\affiliation{Experimentelle Physik 2, Technische Universit\"at Dortmund, 44221 Dortmund, Germany}
\affiliation{Ioffe Physical-Technical Institute, Russian Academy of Sciences, 194021 St.~Petersburg, Russia}
\author{D.~R.~Yakovlev}
\affiliation{Experimentelle Physik 2, Technische Universit\"at Dortmund, 44221 Dortmund, Germany}
\affiliation{Ioffe Physical-Technical Institute, Russian Academy of Sciences, 194021 St.~Petersburg, Russia}
\author{G.~Karczewski}
\affiliation{Institute of Physics, Polish Academy of Sciences, PL-02668 Warsaw, Poland}
\author{M.~Wiater}
\affiliation{Institute of Physics, Polish Academy of Sciences, PL-02668 Warsaw, Poland}
\author{T.~Wojtowicz}
\affiliation{Institute of Physics, Polish Academy of Sciences, PL-02668 Warsaw, Poland}
\affiliation{International Research Centre MagTop, Institute of Physics, Polish Academy of Sciences, PL-02668 Warsaw, Poland}
\author{M.~Bayer}
\affiliation{Experimentelle Physik 2, Technische Universit\"at Dortmund, 44221 Dortmund, Germany}
\affiliation{Ioffe Physical-Technical Institute, Russian Academy of Sciences, 194021 St.~Petersburg, Russia}

\date{\today}

\begin{abstract}

Coherent optical spectroscopy such as four-wave mixing and photon echo generation deliver detailed information on the energy levels involved in optical transitions through the analysis of polarization of the coherent response. In semiconductors, it can be applied to distinguish between different exciton complexes, which is a highly non-trivial problem in optical spectroscopy. We develop a simple approach based on photon echo polarimetry, in which polar plots of the photon echo amplitude are measured as function of the angle $\varphi$ between the linear polarizations of the two exciting pulses. The rosette-like polar plots reveal a distinct difference between the neutral and charged exciton (trion) optical transitions in semiconductor nanostructures. We demonstrate this experimentally by photon echo polarimetry of a 20-nm-thick CdTe/(Cd,Mg)Te quantum well at temperature of 1.5~K. Applying narrow-band optical excitation we selectively excite different exciton complexes including the exciton, the trion, and the donor-bound exciton D$^0$X. We find that polarimetry of the photon echo on the trion and D$^0$X is substantially different from the exciton: The echoes of the trion and D$^0$X are linearly polarized at the angle $2\varphi$ with respect to the first pulse polarization and their amplitudes are weakly dependent on $\varphi$. While on the exciton the photon echo is co-polarized with the second exciting pulse and its amplitude scales as $\cos\varphi$.

\end{abstract}

\maketitle

\section{Introduction}

Four-wave mixing (FWM) spectroscopy provides precise and distinct responses for the different energy level schemes of electronic systems. This was originally demonstrated in gases for different atomic transitions \cite{GordonPR1969, Alekseev1969}. Being applied to semiconductor nanostructures FWM delivers detailed information on the coherent carrier and exciton dynamics \cite{ShahBook}. 

In order to address and manipulate particular optical transitions in semiconductor nanostructures, the underlying exciton complexes such as neutral excitons, charged excitons, bound excitons, biexcitons, etc. should be first identified. Often this is not easy to accomplish when the spectrum is
contributed by several exciton complexes that may spectrally overlap so that their identification is difficult. Even when they can be spectrally separated, their assignment often is complicated as it requires detailed knowledge of the complexes binding energies that are often not known sufficiently accurate. The typical approach then is to employ magneto-optical methods such as detection of polarized photoluminescence, requiring application of strong external magnetic fields \cite{AstakhovPRB1999, GlasbergPRB1999, BayerPRB2002, AkimovAPL2002}. Using polarization-dependent FWM, and in particular photon echo (PE) spectroscopy, makes it possible to perform such an identification without applying the magnetic field.

In semiconductor nanostructures, FWM and PE techniques involving laser pulse sequences with precisely controlled polarizations can be efficiently used as tool to study different exciton complexes such as neutral excitons \cite{SchneiderPRB1994}, charged excitons (trions) \cite{WagnerPRB1999_1, MoodyPRL2014}, and biexcitons \cite{WagnerPRB1999_2, LangbeinPSSa2002}. These techniques have been applied to investigate exciton localization \cite{Steel2002}, many-body interactions \cite{MayerPRB1995, Paul1996, SinghPRB2016}, and excitation-induced dephasing of excitons \cite{Borri1997}. Various protocols of polarized excitation have been used in order to study the coherence of spectrally overlapping exciton states, positively charged trions, and biexcitons in an ensemble of InAs/GaAs quantum dots by two-dimensional Fourier-transform spectroscopy \cite{MoodyPRB2013}.

Polarization-dependent FWM on excitons in semiconductor nanostructures has been subject of extensive research for more than twenty years, however, the majority of studies has been performed in GaAs-based systems, such as quantum wells (QWs) \cite{MayerPRB1995, Paul1996, Steel2002, SinghPRB2016}. In order to understand the complex coherent behavior of excitons localized in GaAs/(Al,Ga)As QWs various nontrivial energy level schemes have been suggested, involving complex selection rules \cite{BottPRB1993, SinghPRB2016, LangbeinPSSa2002}. Moreover, application of spectrally broad femtosecond laser pulses, as often used in FWM spectroscopy, results in excitation of multiple optical states. This unavoidably causes complex many-body interactions affecting the optical selection rules and causing excitation-induced dephasing that shortens the coherent dynamics of the studied optical states.

In this paper, we exploit four-wave mixing spectroscopy with polarization sensitivity to study the polarimetry of photon echoes detected from different exciton complexes in a semiconductor quantum well. We take advantage of picosecond laser pulses to selectively excite individual optical states in a CdTe/(Cd,Mg)Te single QW including the neutral exciton (X), the negative trion (T), and the donor-bound exciton (D$^0$X). This avoids simultaneous generation of multiple optical excitations, which strongly reduces the efficiency of many-body interactions. Operating in the weak excitation regime, we find different polarization properties for the photon echoes from the neutral and charged excitons. Additionally, we perform polarimetry of the photon echoes on D$^0$X, which has not been studied so far with polarization sensitivity. We approve that D$^0$X exhibits a photon echo polarimetry that is in full accordance with the trion model.

\section{Theoretical model}
\label{sec:model}

\begin{figure}[ht]
	\includegraphics[width=\linewidth]{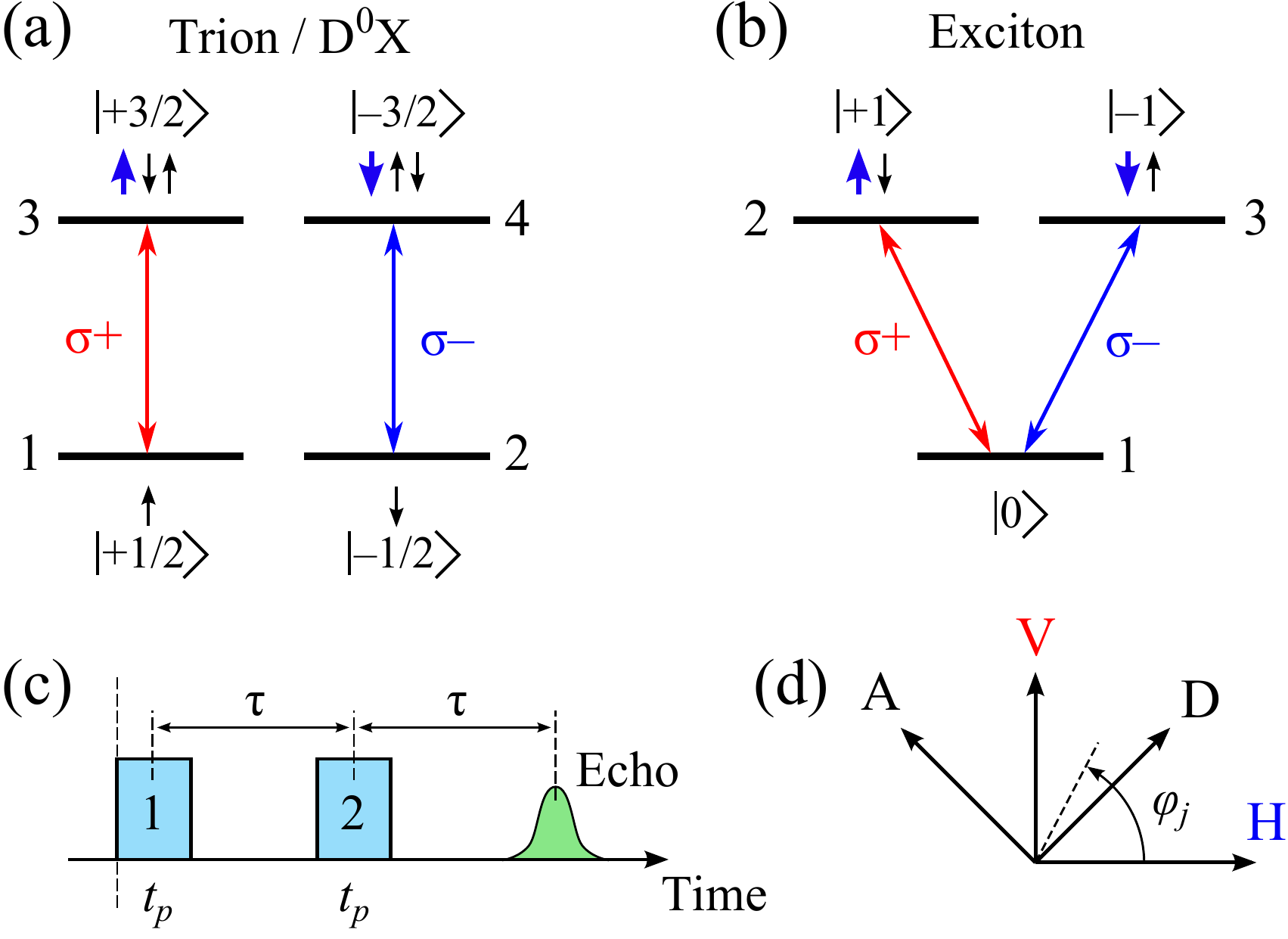}
	\caption{Typical schemes of energy levels and allowed optical transitions in direct band-gap semiconductor structures for the trion/D$^0$X (a) and the exciton (b). Small thin and thick (blue) arrows indicate the spin projections of electrons and holes, respectively; numbers in brackets correspond to the total angular momentum projection values. (c) Temporal sequence of the optical excitation pulses used in the model. (d) Main used polarization directions of excitation pulses and detection: angle $\varphi_j = 0^\circ$, $90^\circ$, $45^\circ$, and $135^\circ (j=1,2,d)$ for horizontal (H), vertical (V), diagonal (D), and anti-diagonal (A) polarizations, respectively.}
	\label{schemes}
\end{figure}

First, we consider the expected polarization properties of the photon echoes on the trion and exciton in a simple theoretical model. Their typical energy schemes with the allowed optical transitions are shown in Fig.~\ref{schemes}. The negative trion scheme consists of a degenerate pair of levels with resident electron spin projections $\ket{\pm1/2}$ and another pair of levels with trion spin projections $\ket{\pm3/2}$ in the ground and excited states of the crystal, respectively. The neutral donor-bound exciton (D$^0$X) is formed from an electron bound to a donor in the ground state and can be described by a similar level scheme. 

In case of the exciton, we consider a simple V-type three-level energy scheme implying that the biexciton binding energy is sufficiently large (about 4.5 meV \cite{Mino2007}), so that resonant excitation of the exciton does not drive the biexciton transition. Such a scheme has zero angular momentum projection in the crystal ground state and $\ket{\pm1}$ in the excited state, as shown in Fig.~\ref{schemes}(b). Due to the optical selections rules, only the optical transitions with difference $\pm1$ in angular momentum projection are allowed, as indicated by the red and blue arrows corresponding to $\sigma_+$ and $\sigma_-$ photons, respectively.

For both cases we compose the Hamiltonian applying the Rotating-Wave Approximation (RWA), $\widehat{H}$, and solve the von Neumann equation to find the evolution of the density matrix $\rho$ in the photon echo experiment:

\begin{equation}
	\label{Neumann}
	i \hbar \dot{\rho} = \left[\widehat{H},\rho \right], \hspace{5mm} \widehat{H} = \widehat{H}_0 + \widehat{V}.
\end{equation}

\noindent We consider the interaction with light $\widehat{V}$ as perturbation of the unperturbed Hamiltonian $\widehat{H}_0$:

\begin{equation*}
\hat V(t) = -\int [\widehat{d}_+(\bm{r}) E_{\sigma_+}(\bm{r},t) + \widehat{d}_-(\bm{r})E_{\sigma_-}(\bm{r},t)] \mathrm d^3 r.
\end{equation*}

\begin{figure*}[ht]
	\centering
	\includegraphics{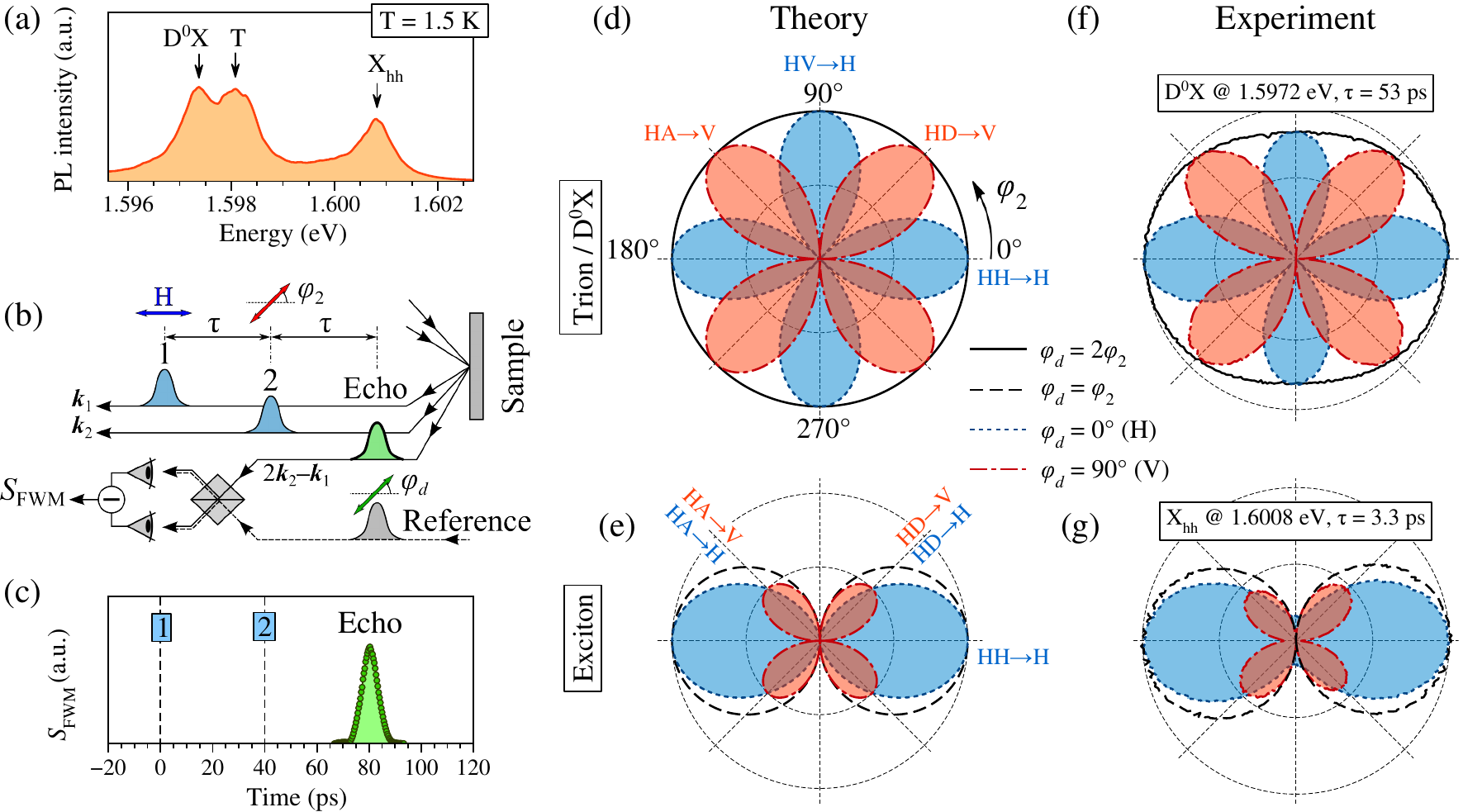}
	\caption{(a) Photoluminescence spectrum of a 20~nm CdTe/(Cd,Mg)Te single QW measured at $T=1.5$~K. (b) Optical scheme of PE detection. (c) PE transient detected from the trion at energy of 1.5981 eV and delay $\tau=40$~ps. (d), (e) Polar rosette-like patterns confirming the theoretically expected behavior of the PE polarization in case of the trion (or D$^0$X) and the exciton, respectively. (f), (g) Polar rosettes of experimentally measured PEs from the D$^0$X at 1.5972~eV, $\tau=53$~ps and exciton at 1.6008~eV, $\tau=3.3$~ps, respectively. In panels (d)-(g) the first pulse is H-polarized ($\varphi_1=0^\circ$); black solid and dashed lines correspond to the polarization of the reference pulse tilted by $\varphi_d=2\varphi_2$ and $\varphi_2$, respectively; blue dotted and red dash-dotted lines correspond to the H and V polarizations of the reference pulse, respectively. The amplitude of the PE is measured in arbitrary units.}
	\label{rosettes}
\end{figure*}

\noindent Here $\widehat{d}_\pm$ are the circularly polarized components of the dipole moment density operator responsible for $\sigma_\pm$ optical transitions and $E_{\sigma_\pm}(\boldsymbol{r},t)$ are those of the electric field of a quasi-monochromatic electromagnetic wave. For the sake of simplicity, we consider rectangular-shaped pulses of duration $t_p$ resulting in the pulse areas $f_{\pm}t_p$ for the two circularly polarized components. Here the $f_{\pm}$ contain smooth envelopes of the circularly polarized components $E_{\sigma_{\pm}}$ and $\hat d_{\pm}$:

\begin{equation} 
\label{circular_excitation}
f_{\pm}(t) = -\frac{2\mathrm e^{i\omega t}}{\hbar}\int \widehat{d}_\pm(\bm r) E_{\sigma_{\pm}}(\bm r,t)\mathrm d^3 \bm r\:.
\end{equation}

\noindent We assume a quasi-linear in power excitation ($\chi^{(3)}$) regime and neglect relaxation.

\subsection{Negatively Charged Trion}

For the trion we compose the following stationary Hamiltonian in RWA with the energy level numbering in accord with Fig.~\ref{schemes}(a) and imply resonant optical excitation corresponding to the Hahn echo case:

\begin{align*}
	\widehat{H} =
	\begin{pmatrix}
		0 & 0 & f_+^* & 0 \\
		0 & 0 & 0 & f_-^* \\
		f_+ & 0 & \Delta & 0\\
		0 & f_- & 0 & \Delta
	\end{pmatrix},
\end{align*}
\noindent where $\Delta$ is the oscillator detuning.

We solve Eq.~(\ref{Neumann}) for the following sequence of operations on the system, as shown in Fig.~\ref{schemes}(c): The system is excited by the first laser pulse during time interval $t_p$ with constant circular components $f_{1\pm}$. Then the system experiences free evolution during the time interval $\tau-t_p$, which is followed by the action of the second laser pulse with components $f_{2\pm}$. After the second period of free evolution during the time interval $\tau-t_p/2$ the oscillators rephase and the photon echo polarization builds up, which is analyzed in a specific polarization. Considering the initial density matrix $\rho^0$ with the only nonzero terms $\rho^0_{11}=\rho^0_{22}=1/2$, we find the following analytical solution for the circular components of the PE field from the trion:

\begin{align}
	\label{result_trion}
	P_{\sigma_+}^T &\propto t_p^3f_{1+}\left(f_{2+}^*\right)^2, \\
	P_{\sigma_-}^T &\propto t_p^3f_{1-}\left(f_{2-}^*\right)^2.	\nonumber
\end{align}

\subsection{Exciton}

We perform the equivalent procedure for the exciton. The initial 3$\times$3 density matrix has only the ground state populated, $\rho^0_{11}=1$, and the Hamiltonian reads as follows:

\begin{align*}
	\widehat{H} =
	\begin{pmatrix}
		0 & f_+^* & f_-^* \\
		f_+ & \Delta & 0 \\
		f_- & 0 & \Delta
	\end{pmatrix}.
\end{align*}

\noindent The analytical solution for the PE from the exciton gives us the following circular components:

\begin{align}
	\label{result_exciton}
	P_{\sigma_+}^X &\propto t_p^3f_{2+}^*\left(f_{2+}^*f_{1+}+f_{2-}^*f_{1-}\right), \\
	P_{\sigma_-}^X &\propto t_p^3f_{2-}^*\left(f_{2-}^*f_{1-}+f_{2+}^*f_{1+}\right).	\nonumber
\end{align}

\subsection{Method}

In the photon echo polarimetry, which we develop here, two linearly polarized exciting pulses are used to generate the photon echo. The angle between the linear polarizations of the exciting pulses, $\varphi$, is varied and the polarization of the PE is analyzed. Namely, we analyze the PE polarization by probing it in a certain linear polarization (detection), as described in detail below.

To specify a certain polarization scheme for the excitation and detection we use the notation in form: HV$\rightarrow$H. Here, the polarizations of the two exciting pulses and of the analyzed one are indicated correspondingly before and after the arrow. The symbols H, V, D, and A correspond to the polarization angles $\varphi_j= 0^\circ$, $90^\circ$, $45^\circ$, and $135^\circ (j=1,2,d)$, as indicated in Fig.~\ref{schemes}(d).

Now we use Eqs.~(\ref{circular_excitation})-(\ref{result_exciton}) to calculate the expected PE polarization from the trion (D$^0$X) and the exciton generated by the two linearly polarized pulses. The linearly polarized excitation of the $j$-th exciting pulse ($j=1,2$) with amplitude $E_j$ and tilt angle $\varphi_j$ can be expressed in the following way:

\begin{equation} 
	\label{polarizations}
	\begin{pmatrix} 
		E^j_{\sigma_+} \\ 
		E^j_{\sigma_-} 
	\end{pmatrix} = \frac{1}{\sqrt{2}}E_j
	\begin{pmatrix} 
		\mathrm e^{i\varphi_j}\\ 
		\mathrm e^{-i\varphi_j}
	\end{pmatrix}.
\end{equation}  \\

\noindent Accordingly, the modulus of the linearly polarized component of the detected PE amplitude $P_d$, defined by the tilt angle $\varphi_d$, for the optical transition $k$ ($k=X$ or $T$), reads as

\begin{equation}
\label{detection}
P_d = \left|\mathrm e^{-i\varphi_d}P_{\sigma_+}^k + \mathrm e^{i\varphi_d}P_{\sigma_-}^k\right|.
\end{equation}

It is worth noting that any kind of FWM response measured in the $\chi^{(3)}$-regime including the free-polarization decay (self-diffraction of the second pulse) or the photon echo should obey the same polarization rules as described by Eqs.~(\ref{result_trion})--(\ref{detection}).

The polarized PE when the first pulse is H-polarized and the $\varphi_2$ angle is varied, calculated using Eqs.~(\ref{result_trion})--(\ref{detection}), is shown in Fig.~\ref{rosettes}(d) and \ref{rosettes}(e) for the trion (D$^0$X) and the exciton, respectively. For brevity, due to their characteristic shape the polar diagrams resulting from this analysis will be called polar rosettes in the following.

In the considered excitation sequence, the trion (D$^0$X) PE is linearly polarized with the angle $\varphi_\text{PE}=2\varphi_2$ and the PE amplitude is independent of $\varphi_2$. This is shown in Fig.~\ref{rosettes}(d) with the solid line circle. Detection of the H-polarized component results in the photon echo PE$_{\text{H}}^T(\varphi_2)\propto|\cos2\varphi_2|$, which produces four maxima in the configurations HH$\rightarrow$H and HV$\rightarrow$H (the blue dot line). Correspondingly, detection of the V-polarized component gives PE$_{\text{V}}^T(\varphi_2)\propto|\sin2\varphi_2|$ with four maxima in the HD$\rightarrow$V and HA$\rightarrow$V configurations. 

\begin{figure*}[t]
	\centering
	\includegraphics{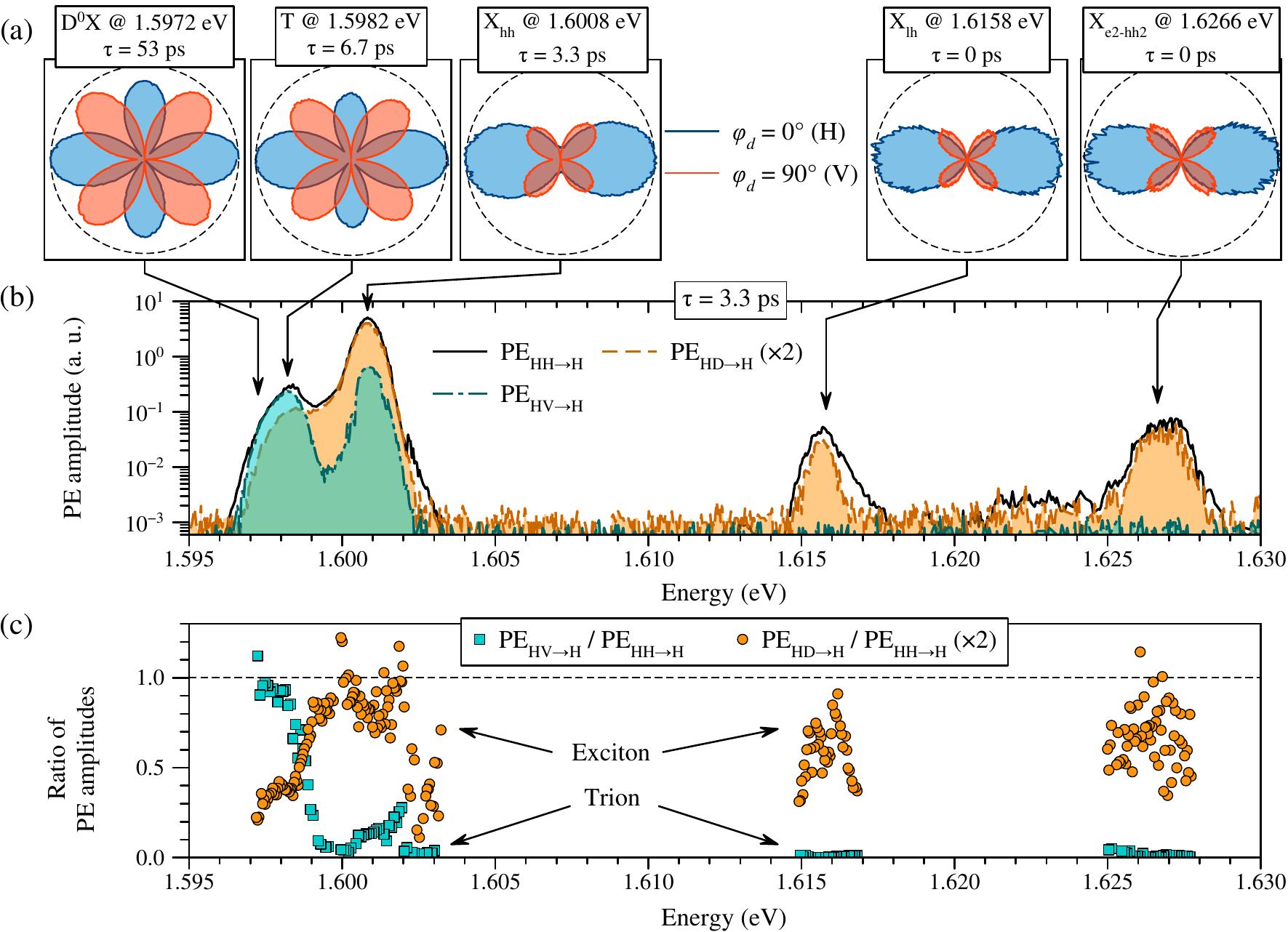}
	\caption{Results of polarized photon echo measurements from the CdTe/(Cd,Mg)Te single QW carried out with the spectroscopic resolution: (a) Polar diagrams displaying the dependence of PE amplitude on the second pulse polarization angle $\varphi_2$ measured at two angles of the reference pulse, $\varphi_d = 0^\circ$ (blue line) and $90^\circ$ (red line), from the donor-bound exciton D$^0$X, the trion T, the heavy hole exciton X$_\text{hh}$, the light hole exciton X$_\text{lh}$, and the $1S$ state of the X$_\text{e2-hh2}$ exciton. Excitation energy and pulse delay $\tau$ are indicated in the legends. Scales in the panels are normalized by the PE$_{\text{HH}\rightarrow\text{H}}$ amplitude. (b) Spectral dependences of the PE amplitude measured at $\tau=3.3$~ps in the following polarization configurations: HH$\rightarrow$H (black solid line), HV$\rightarrow$H (cyan dash-dotted line), and HD$\rightarrow$H (orange dashed line) scaled twice. (c) Ratios of the PE amplitudes, which express the trion (cyan squares) and exciton (orange circles) contributions to the polarimetric PE signal.}
	\label{spectra}
\end{figure*}

The same excitation sequence applied to the exciton should result in the photon echo being co-polarized with the second pulse ($\varphi_\text{PE}=\varphi_2$) with the PE amplitude $\propto|\cos\varphi_2|$. This is shown by the two black dashed circles in Fig.~\ref{rosettes}(e). The H-component of the exciton photon echo PE$_{\text{H}}^X(\varphi_2)\propto\cos^2\varphi_2$ displays only two maxima in the HH$\rightarrow$H configuration and zero signal in the HV$\rightarrow$H configuration. The four-lobe V-component of the exciton photon echo PE$_{\text{V}}^X(\varphi_2)\propto|\sin2\varphi_2|/2$ is similar to that of the trion PE, but twice smaller in magnitude.

\section{Sample and Experimental technique}
\label{sec:sample}

In order to demonstrate the potential of photon echo polarimetry we choose a 20-nm-thick CdTe/Cd$_{0.76}$Mg$_{0.24}$Te single QW grown by the molecular-beam epitaxy (sample 032112B). Details of the structure growth can be found in Ref.~\cite{PoltavtsevCdTe2017}. The structure was not intentionally doped with donors and the resident electron density $n_d\leqslant10^{10}$~cm$^{2}$ is due to an unavoidable background of impurities. Illumination of the sample by a weak unfocused green cw-laser above barrier (photon energy 2.33~eV) was used to generate photo-injected carriers, which results in an increased population density of trions. Figure~\ref{rosettes}(a) displays the photoluminescence (PL) spectrum of the studied QW measured at the temperature of 1.5~K. It manifests lines of the heavy hole exciton X$_\text{hh}$, the negatively charged exciton (trion) T located 2.8~meV below, and the neutral donor-bound exciton D$^0$X split off from the trion by 0.6~meV.

To generate photon echoes and study the PE polarimetry we implement a time-resolved four-wave mixing technique \cite{SemicondBook2008}. The sample mounted in a helium bath cryostat and cooled down to 1.5~K is excited by a sequence of two picosecond laser pulses generated by a tunable Ti:sapphire laser. The first and second pulses with wavevectors $\bm{k}_1$ and $\bm{k}_2$, respectively, are focused into a spot of about 200~$\mu$m in diameter. Their directions are close to the sample normal. The pulses are separated in time by a variable delay $\tau$ tuned by means of an optical delay line. The laser pulse duration is about 2~ps. The FWM signal is collected in the reflection with wavevector $2\bm{k}_2-\bm{k}_1$ and mixed with an additional reference pulse on the photodetector, as shown in Fig.~\ref{rosettes}(b). This optical heterodyning is applied in order to accomplish a background-free high-sensitivity measurement of the FWM signal. The modulus of the cross-correlation of the FWM signal amplitude $E_\text{FWM}$ with the reference pulse field $E_\text{Ref}$ is detected at the output of the photodetector: $S_\text{FWM}\propto|E_\text{FWM}E_\text{Ref}^*|$. Thus, only the polarization component of the FWM signal defined by the reference pulse is measured. A detailed description of this technique can be found in Ref.~\cite{CdTeArSSP2018}.

Figure~\ref{rosettes}(c) demonstrates a typical FWM transient measured at the trion resonance (1.5981~eV) for a delay $\tau=40$~ps. The observed coherent burst occurring at $2\tau=80$~ps delay is the PE pulse convoluted with the reference pulse. In the following we study the PE amplitude when the reference pulse is centered exactly at $2\tau$ delay.

It is known from previous studies carried out on the same QW structure that optical excitation at elevated power densities may result in coherent Rabi flops of the PE amplitude accompanied by temporal shifts of the echo pulse, which were recently studied using the time-resolved FWM technique \cite{PoltavtsevCdTe2017}. In order to avoid this effect and stay in the quasi-linear $\chi^{(3)}$-regime we employ low-power pulses providing a total pulse power below 20~nJ/cm$^2$. The X$_\text{hh}$, T, and D$^0$X optical transitions in this QW were also studied with temporal and spectroscopic resolutions, and the spectral dependence of the coherence time $T_2$ was measured \cite{PoltavtsevCdTe2017}. Here we focus mostly on the polarization properties of the photon echoes from these transitions.

\section{Experimental results and discussion}
\label{sec:results}

In order to reveal differences in the polarization properties of the photon echoes on the trion (D$^0$X) and the exciton we excite the QW with an H-polarized first pulse ($\varphi_1=0^\circ$) and vary the polarization angle of the second pulse $\varphi_2=0...360^\circ$. The detected PE is analyzed at $\varphi_d=0^\circ$ (H) and $90^\circ$ (V) by a proper choice of the reference pulse polarization. Additionally, the polarization of the detection is varied synchronously with the second pulse polarization, $\varphi_d=2\varphi_2$ or $\varphi_d=\varphi_2$. This allows us to measure the photon echo amplitude in its expected linear polarization both from the trion (D$^0$X) and the exciton, respectively.

\subsection{Polarimetry of photon echo}

Figures~\ref{rosettes}(f) and \ref{rosettes}(g) display polar rosettes of the polarimetric echo signals measured on the D$^0$X at energy of 1.5972~eV and on the exciton at energy of 1.6008~eV. The pulse delays used in these measurements are $\tau=53$~ps and 3.3~ps. The measurements on both systems are in good agreement with the theoretical expectations. There is, however, some reduction of the PE signal measured on D$^0$X in the HV$\rightarrow$H polarization configuration, which appears as a squeezing of the trion (D$^0$X) polar rosette along the V-direction.

From these rosettes we can extract distinct differences, which can be used to identify the trion and the exciton contributions to the PE polarimetric signal: In the HV$\rightarrow$H polarization configuration the trion has a strong PE signal, while the exciton exhibits almost zero signal. In the HD$\rightarrow$H configuration, on the other hand, the exciton provides half of the maximal PE amplitude (measured in HH$\rightarrow$H), while the trion manifests no PE signal.

Using these characteristics we accomplish photon echo polarimetry with spectroscopic resolution aimed at observing separately the trion and exciton contributions to the echo signal. To carry out such a separation of contributions we tune the laser energy in the vicinity of optical transitions present and detect the PE amplitude in different polarization configurations at a relatively short pulse delay ($\tau=3.3$~ps). These spectroscopic measurements are presented in Fig.~\ref{spectra}(b). The PE amplitude in the HH$\rightarrow$H configuration (PE$_{\text{HH}\rightarrow\text{H}}$, black solid line), which is not sensitive to the type of the optical transition, exhibits five resonant lines: Additionally to the donor-bound exciton D$^0$X (1.5972~eV), the trion T (1.5982~eV), and the heavy hole exciton X$_\text{hh}$ (1.6008~eV) present also in the PL spectrum [Fig.~\ref{rosettes}(a)] the spectral lines of the light hole exciton at 1.6158~eV and the $1S$ state of the X$_\text{e2-hh2}$ exciton at 1.6266~eV are observed. All features are broadened due to convolution with the laser pulse with a full width at half-maximum of 0.9~meV.

Detection in the HV$\rightarrow$H configuration results in a substantially different spectrum (PE$_{\text{HV}\rightarrow\text{H}}$ in Fig.~\ref{spectra}(b), cyan dash-dotted line), which deviates strongly from the PE$_{\text{HH}\rightarrow\text{H}}$ spectrum in the vicinity of the heavy hole exciton and manifests zero signal around the light hole exciton and the X$_\text{e2-hh2}$ exciton. 

The PE spectrum measured in the HD$\rightarrow$H configuration (PE$_{\text{HD}\rightarrow\text{H}}$) is plotted in Fig.~\ref{spectra}(b) scaled twofold with the orange dashed line. By the contrast, it correlates strongly with the PE$_{\text{HH}\rightarrow\text{H}}$ data in the vicinity of the X$_\text{hh}$, X$_\text{lh}$, and X$_\text{e2-hh2}$ exciton transitions and is substantially reduced around the D$^0$X and trion resonances. 

Figure~\ref{spectra}(c) displays the spectral dependences of both the trion (squares) and exciton (circles) contributions to the PE signal obtained as the signal ratios PE$_{\text{HV}\rightarrow\text{H}} / $PE$_{\text{HH}\rightarrow\text{H}}$ and 2PE$_{\text{HD}\rightarrow\text{H}} / $PE$_{\text{HH}\rightarrow\text{H}}$, respectively. In the low-energy range these spectra clearly demonstrate a separation of the two different types of optical transitions. The echo signal measured below 1.5988~eV manifests mostly trion contributions, however, some contribution of the exciton signal is also observed. This effective exciton contribution is associated with the reduction of the trion echo signal in the HV$\rightarrow$H polarization configuration, shown before. It is remarkable that both light hole X$_\text{lh}$ and X$_\text{e2-hh2}$ exciton features that are far from any trion transitions display pure excitonic character (PE$_{\text{HV}\rightarrow\text{H}}=0$).

The polar rosettes measured by the procedure described above on every of the optical transitions are shown in the panels of Fig.~\ref{spectra}(a). One can see from these polar rosettes that the trion excited at the energy of 1.5982~eV exhibits an even stronger reduction of the PE$_{\text{HV}\rightarrow\text{H}}$ component than the D$^0$X excited at 1.5972~eV. The polar rosettes measured on the X$_\text{lh}$ and X$_\text{e2-hh2}$ excitons are similar to those obtained by the theoretical model for the neutral exciton [Fig.~\ref{rosettes}(e)].

\begin{figure}[t]
	\centering
	\includegraphics{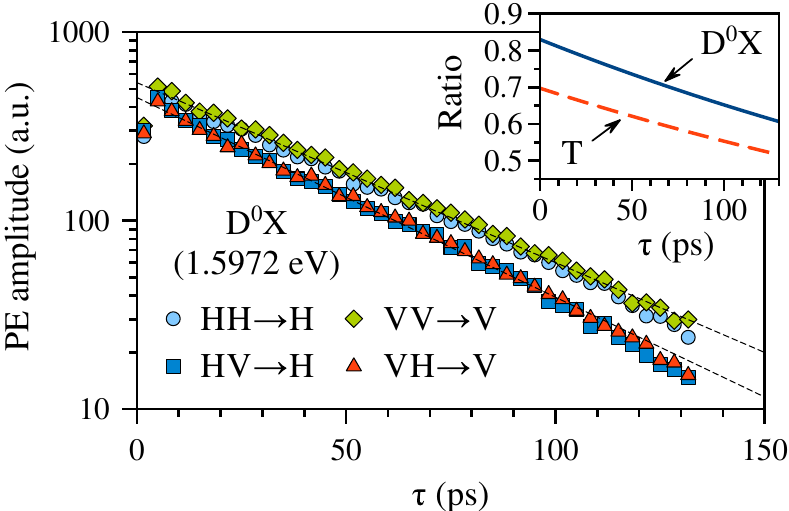}
	\caption{Photon echo decays measured on D$^0$X (1.5972~eV) as dependences of the PE amplitude on the pulse delay $\tau$ for various polarization configurations (symbols). Fits by $B\exp(-2\tau / T_2)$ are given with dashed lines: for HH$\rightarrow$H (VV$\rightarrow$V) $B=540, T_2=91$~ps; for HV$\rightarrow$H (VH$\rightarrow$V) $B=450, T_2=82$~ps. Inset shows the ratios PE$_{\text{HV}\rightarrow\text{H}} / $PE$_{\text{HH}\rightarrow\text{H}}$ of the two fit curves for D$^0$X (1.5972~eV) by the solid line and for the trion (1.5982~eV) by the dashed line.}
	\label{D0X_decays}
\end{figure}

In order to understand how the ratio PE$_{\text{HV}\rightarrow\text{H}} / $PE$_{\text{HH}\rightarrow\text{H}}$ of the trion (D$^0$X) PE signals changes with the pulse delay $\tau$ we measure the exponential PE decays by varying $\tau$ and detecting the PE amplitude for co-polarized and cross-polarized excitation. The decays of the PE signal on D$^0$X, shown in Fig.~\ref{D0X_decays}, manifest slightly different coherence times, $T_2$(HH$\rightarrow$H)=91~ps and $T_2$(HV$\rightarrow$H)=82~ps, and amplitudes. We checked that rotation of the whole polarization basis by $90^\circ$ produces similar results, i.e. $T_2$(VV$\rightarrow$V)=$T_2$(HH$\rightarrow$H) and $T_2$(VH$\rightarrow$V)=$T_2$(HV$\rightarrow$H). Similar measurements on the trion (1.5982~eV) gave $T_2$(HH$\rightarrow$H)=73~ps and $T_2$(HV$\rightarrow$H)=67~ps (decays not shown). The inset in Fig.~\ref{D0X_decays} displays the ratios of the PE decay fits in the cross- and co-polarized configurations for both the D$^0$X and T transitions. We do not discuss the origin of the small coherence time difference here and leave this subject for future studies. We conclude that the observed type of anisotropy does not relate, in lowest approximation, to the crystal axes orientation, but is a fundamental property of the studied optical transitions excited in the co- or cross-polarized configuration.

\subsection{Discussion}

Polarization-dependent FWM and PE generation on excitons in semiconductor QWs have been intensively studied before. Yaffe et al. observed a strong reduction of the exciton echo intensity for cross-polarized excitation of 3-nm-thick GaAs/(Al,Ga)As QWs, although the echo polarization was not analyzed \cite{Yaffe1993}. Schneider and Ploog observed the correspondence of the PE polarization to the second pulse polarization, as well as the absence of the echo signal in cross-polarized excitation of the exciton in a 12-nm-thick GaAs/AlAs QW at low temperature \cite{SchneiderPRB1994}. The complete set of Stokes parameters of the FWM response from the exciton in GaAs/(Al,Ga)As multiple QWs was measured and analyzed with femtosecond temporal resolution by Paul et al. \cite{Paul1996}. They concluded that, in general, the polarization of the FWM signal is highly elliptical and shows a complex temporal dynamics. This was explained by complex many-body interactions and a five-level scheme of the involved optical transitions. The group of Cundiff published several papers on polarization-dependent photon echoes on excitons localized in GaAs/(Al,Ga)As QWs \cite{HuPRB1994, Steel2002, SinghPRB2016}. These studies, however, deal with strong optical excitation causing many-body interactions to strongly affect the coherent dynamics of excitons and the optical selection rules. Wagner et al. employed FWM spectroscopy using femtosecond pulses and different polarization protocols of excitation to distinguish between the trion and the biexciton \cite{WagnerPRB1999_1}, and between the heavy-heavy-hole biexciton and heavy-light-hole biexciton possessing different optical selection rules \cite{WagnerPRB1999_2} in a single ZnSe/(Zn,Mg)Se QW.

In the present paper, we have chosen the model system of a CdTe/(Cd,Mg)Te QW with spectrally well isolated optical transitions, that was previously studied by other FWM-based techniques \cite{PoltavtsevCdTe2017, SalewskiPRX2017}. In particular, to precisely distinguish between the donor-bound exciton and trion transitions three-pulse photon echo spectroscopy in an external magnetic field was carried out \cite{SalewskiPRX2017}. These two optical transitions were distinguished using a difference in the g-factor, whose spectral dependence revealed a step-like behavior in the relevant spectral range. In another FWM study optical Rabi flops could be observed in the photon echo amplitude on the same sample \cite{PoltavtsevCdTe2017}. The flops manifested a substantial difference between the D$^0$X and the trion: Rabi oscillations from D$^0$X experience less damping compared to the trion. Based on the different FWM studies, it can be generally concluded that D$^0$X tends to be substantially more robust in the coherent optical behavior than the trion. In addition to the longer-lived coherent dynamics, here we find that D$^0$X shows a polarimetry of photon echoes that is in a close correspondence to the 4-level trion model. We believe that this is due to stronger localization of D$^0$X than of the trion.

As an extension, the presented method can be used to study other exciton complexes. In particular, the positive trion, as well as the neutral acceptor-bound exciton (A$^0$X), have a similar energy scheme as the negative trion, so that the trion polar rosette is expected to show up in their PE polarimetry. Similarly, the ionized donor-bound exciton (D$^+$X) and the ionized acceptor-bound exciton (A$^-$X) are expected to exhibit exciton-like polar rosettes.

\section{Conclusions}

We have suggested a photon echo polarimetry method to identify the origin of the exciton complexes in semiconductor nanostructures. To prove this method, we adjusted the FWM technique for precise polarimetric measurements by providing sufficient spectroscopic resolution and applied it to the CdTe/(Cd,Mg)Te quantum well. Using certain polarization protocols in optical excitation and analyzing the polarization of the photon echoes we have clearly distinguished between the trion and the exciton. We have shown that the donor-bound exciton, due to its energy level structure, manifests polarization properties of photon echoes qualitatively similar to those of the trion. The neutral excitons generate photon echoes in accordance to a simple V-type system. Deviations of the observed photon echo polarimetry from the presented simple theoretical models may be studied to understand coherent interactions between the different exciton complexes which may cause modifications in the optical selection rules. In particular, the photon echo polarimetry can be applied to materials, in which the exciton complexes are known to have complex optical selection rules or have not been studied well so far. Among them, materials such as halide perovskites \cite{March2016, Bohn2018} or transition metal dichalcogenide monolayers \cite{MoodyNC2015, Hao2016, SinghTMDC2016, JakubczykNL2016, HaoNP2016, DeyPRL2016} can be suggested. Additionally, wurtzite ZnO-based materials are of interest, for which robust coherent dynamics of the different optical transitions were recently observed though photon echoes \cite{PoltavtsevPRB2017, SolovevPRB2018}.

\section*{Acknowledgments}

The authors thank G.~G. Kozlov and M.~M. Glazov for helpful discussions and A.~N. Kosarev for assistance in the experiments.

\section*{Funding}
Deutsche Forschungsgemeinschaft (DFG), Collaborative Research Centre TRR 142 (Project No. A02); DFG, International Collaborative Research Centre 160 (Project No. A3); Russian Foundation for Basic Research (Projects No. 15-52-12016 NNIO\_a and 16-29-03115 ofi\_m); St. Petersburg State University (Projects No. 11.34.2.2012 and Pure-SPBU-34825718).


\end{document}